\documentclass[aps,pra,reprint,twocolumn,showpacs,showkeys]{revtex4-1}
\usepackage{amsmath,amssymb,amstext}
\usepackage[usenames,dvipsnames]{color}
\usepackage{graphicx}
\usepackage{bm,bbold,braket}
\usepackage{natbib}
\usepackage{txfonts, comment,stmaryrd}
\usepackage{dcolumn}
\usepackage{color}
\usepackage{xcolor}
\colorlet{rn}{red}
\colorlet{an}{blue}
\usepackage[english]{babel}
\usepackage{mathcomp}
\usepackage{wasysym}
\usepackage[colorlinks,bookmarks=false,citecolor=blue,linkcolor=red,urlcolor=blue]{hyperref}
\begin{document}

\title{Stripe and checkerboard patterns in a stack of driven quasi-one-dimensional dipolar condensates}
\author{Shreyas Nadiger}
\affiliation{Department of Physics, Indian Institute of Science Education and Research, Pune- 411008, India}
\author{Sandra M. Jose}
\affiliation{Department of Physics, Indian Institute of Science Education and Research, Pune- 411008, India}
\author{Ratheejit Ghosh}
\affiliation{Department of Physics, Indian Institute of Science Education and Research, Pune- 411008, India}
\author{Inderpreet Kaur}
\affiliation{Department of Physics, Indian Institute of Science Education and Research, Pune- 411008, India}
\author{Rejish Nath} 
\affiliation{Department of Physics, Indian Institute of Science Education and Research, Pune- 411008, India}

\begin{abstract}
The emergence of transient checkerboard and stripe patterns in a stack of driven quasi-one-dimensional homogeneous dipolar condensates is studied. The parametric driving of the $s$-wave scattering length leads to the excitation of the lowest collective Bogoliubov mode. The character of the lowest mode depends critically on the orientation of the dipoles, corresponding to out-of-phase and in-phase density modulations in neighboring condensates, resulting in checkerboard and stripe patterns. Further, we show that a dynamical transition between the checkerboard and stripe patterns can be realized by quenching the dipole orientation either linearly or abruptly once the initial pattern is formed via periodic driving.
\end{abstract}



\maketitle

\section{Introduction}
%
The long-range and anisotropic dipole-dipole interactions (DDIs) lead to many phenomena in quantum gases \cite{bar08,lah09, bar12, cho23,def23}, including self-organized, equilibrium density patterns in Bose-Einstein condensates (BECs) \cite{her21}. The patterns also emerge when a dipolar gas is quenched or tuned to instabilities \cite{nat09, mac12, rag15, mis16, kui18, zha22} that can be done by varying the $s$-wave scattering length, the number of particles, or the orientation of dipoles. Additionally, the long-range character of the DDI can lead to collective phenomena among physically disconnected condensates \cite{jun10,ros13}. They include dipolar drag \cite{mat11,gal14}, hybrid excitations \cite{kla09, hua10,mal13}, soliton complexes \cite{nat07,kob09,lak12,lakom12,elh19,heg21}, supersolids \cite{pra23} and coupled density patterns \cite{lako12}. Equilibrium checkerboard and stripe density waves are predicted to exist in disconnected dipolar gases in a planar array of one-dimensional tubes \cite{kol08}. In condensates, it is found that the inter-layer interactions can significantly affect the stability criteria, even in chromium condensates for which the dipole moment is comparatively weak \cite{mul11, wil11}. Remarkably, a recent experiment using cold gas of dysprosium atoms demonstrates strong inter-layer dipolar effects by reducing the layer separations to 50 nm from typical lattice spacings of 500 nm \cite{lid23}, opening up various perspectives in the physics of dipolar gases. Moreover, the recent achievement of ground-state polar molecules \cite{big23} and the successive production of a Bose-Einstein condensate of polar molecules \cite{nic23} also offer promising avenues to explore inter-layer effects.

When condensates are subjected to time-periodic interaction parameters \cite{sta02,nat10, kat10,lako12,ngu19, zha20,kwo21,san23, kei23, nik23} or external potentials \cite{sta04,mod06,eng07,nic07,nic11,bal12,bal14,ver17,com22}, transient density patterns are formed. For small modulation amplitudes, Bogoliubov modes of the initial state govern the periodicity of such patterns via resonant conditions \cite{sta02,eng07,com22}. Pattern formation under periodic driving is explored in a single \cite{nat10, lako12, vud19,tur20} and a pair of quasi-one-dimensional (Q1D) dipolar BECs \cite{lako12}. However, not technically challenging, parametric driving still needs to be experimentally demonstrated in dipolar condensates. In contrast to the condensates with short-range interactions, a rotonic spectrum of a dipolar condensate can lead to unique and non-trivial wavenumber selection, making higher harmonics more unstable than lower ones \cite{nat10, lako12, tur20}. For a pair of Q1D dipolar condensates, the long-range DDIs lead to symmetric and antisymmetric modes, which result in transient symmetric and antisymmetric Faraday patterns under periodic driving \cite{lako12}. 

\begin{figure} 
\centering
    \includegraphics[width=\linewidth]{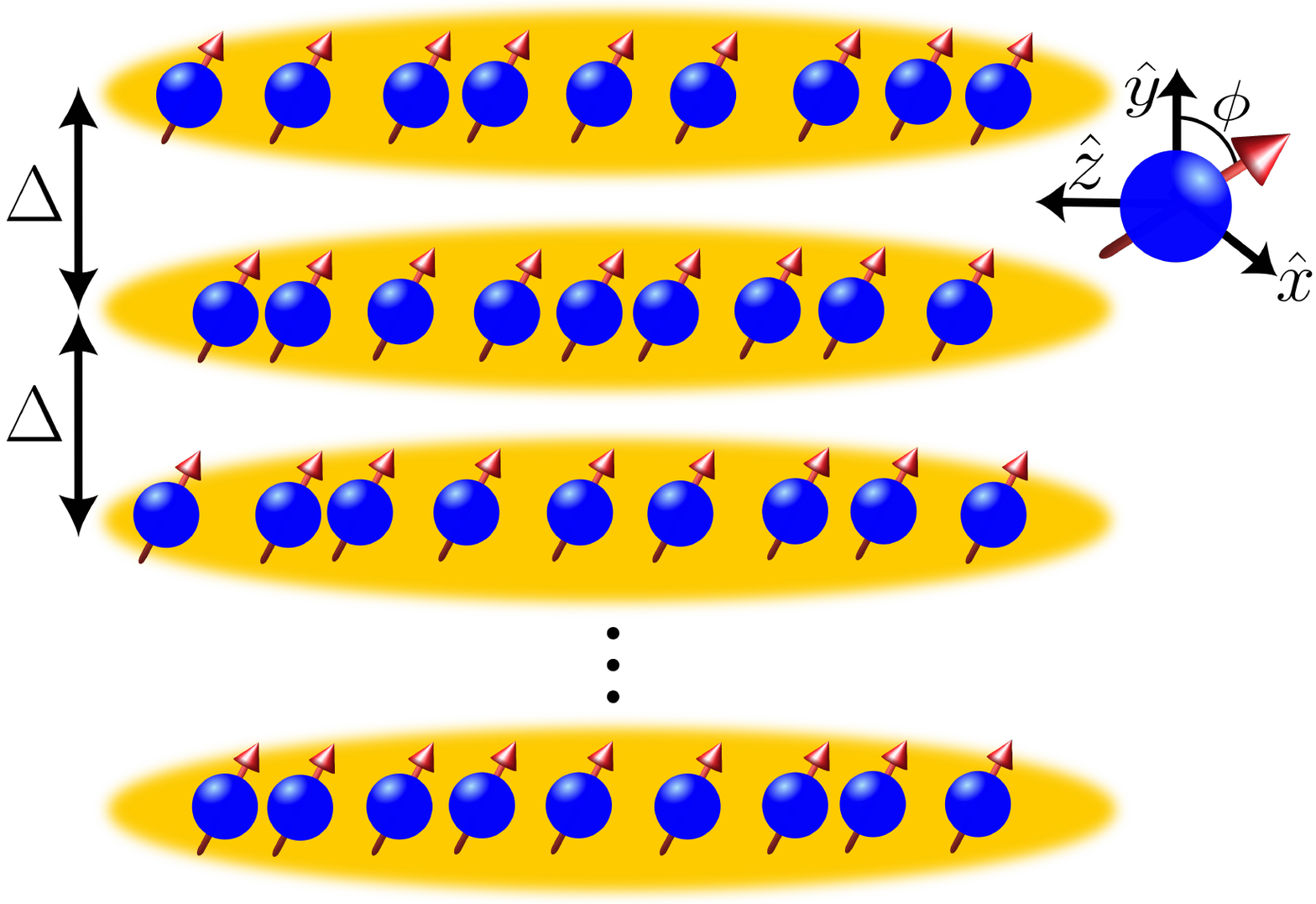}
    \caption{The schematic setup of a stack of Q1D dipolar BECs. The angle $\phi$, between the dipole moment and the $y$-axis (in the plane), determines the nature of inter-tube DDIs, and $\Delta$ is the separation between the adjacent Q1D condensates along the $y$-axis.}
    \label{fig:1}
\end{figure}

In this paper, we show that by engineering the Bogoliubov modes via dipole orientation, transient stripe, and checkerboard patterns can be created in a stack of driven Q1D homogeneous dipolar condensates. We consider the parametric driving of the $s$-wave scattering length, which excites the lowest branch of the Bogoliubov spectrum. When the dipoles are aligned such that the inter-tube DDIs are repulsive, the lowest mode describes out-of-phase density modulations in neighboring condensates, leading to a checkerboard pattern. Making the inter-tube DDIs attractive, the nature of the lowest mode changes to in-phase density modulations, causing stripe patterns. Density-density correlations between the condensates in different tubes can distinguish the two patterns. Finally, we show that once the initial pattern is formed via periodic driving, quenching the dipole orientation leads to a dynamical transition between the checkerboard and stripe patterns. 

The paper is structured as follows. In Sec.~\ref{setup} we introduce the setup of a stack of Q1D dipolar homogeneous BECs, the governing equations and the Bogoliubov spectrum. The Mathieu equations describing the driven condensates are discussed in Sec. \ref{pd}. The formation of stripe and checkerboard patterns are discussed in Secs.~\ref{sp} and \ref{cbp}, respectively.  The effect of quenching of the dipole orientation on the initial pattern is analyzed in Sec.~\ref{qda}. Finally, we summarize and provide an outlook in Sec.~\ref{sum}.
\section{BEC setup and Bogoliubov spectrum}
\label{setup}
We consider a stack of $N$ homogeneous, Q1D dipolar BECs as shown in Fig.~\ref{fig:1}. The condensates form an array along the $y$-axis with a distance of $\Delta$ between the adjacent ones. $\Delta$ is taken sufficiently large to prevent overlap between the condensates. The dipoles are polarized in the $ xy$ plane, making an angle $\phi$ with the $y$-axis. The intra-tube DDIs are repulsive irrespective of the value of $\phi$, whereas the inter-tube DDIs are attractive for $\phi=0$ and are repulsive when $\phi=$90 degrees. At very low temperatures, the system is described by coupled non-local Gross-Pitaevskii equations (NLGPEs),
\begin{align}
    i\hbar\frac{\partial\psi_j(z)}{\partial t}=\bigg[ -\frac{\hbar^2}{2M}\frac{\partial^2}{\partial z^2} + gn_j(z)+ \frac{g_d}{3} \sum_{l=1}^N \int \frac{dq}{2\pi} e^{iqz} n_l(q) \nonumber \\
 \times F_{|l-j|}(q) \bigg] \psi_j(z),
   \label{gpe}
\end{align}
where $\psi_j(z)$ is the wave function of the $j$th condensate. The parameter $g=2\hbar^2a_s/ml_\perp^2$ quantifies strength of short-range interactions with $a_s$ being the s-wave scattering length and $l_{\perp} = \sqrt{\hbar/m\omega_{\perp}}$ where $\omega_{\perp}$ is the transverse trap frequency.  $g_d\propto d^2$ provides the strength of the DDIs, $n_l(q)$ is the Fourier transform of the condensate density $n_l(z) = |\psi_l(z)|^2$ and the function,
\begin{eqnarray} \label{eq:F_m(k)}
    F_{j}(q) =  \int_0^{\infty} dk\frac{ke^{-\frac{1}{2}k^2 l_{\perp}^2}}{k^2 + q^2}\left[ (k^2 - 2q^2)J_0(j\*k\*\Delta)\right. \nonumber\\
     \left. - 3k^2 \cos{(2\phi)} \, J_2(j\*k\*\Delta) \right],
     \label{fn}
\end{eqnarray}
depends on $\phi$ and $\Delta$, where $J_n(x)$ is the Bessel function of the first kind. The Q1D nature of each condensate demands $\mu_j\ll\hbar\omega_\perp$, $\forall j$ where $\mu_j$ is the chemical potential of the $j$th condensate.

\begin{figure} 
        \centering
        \includegraphics[width=\linewidth]{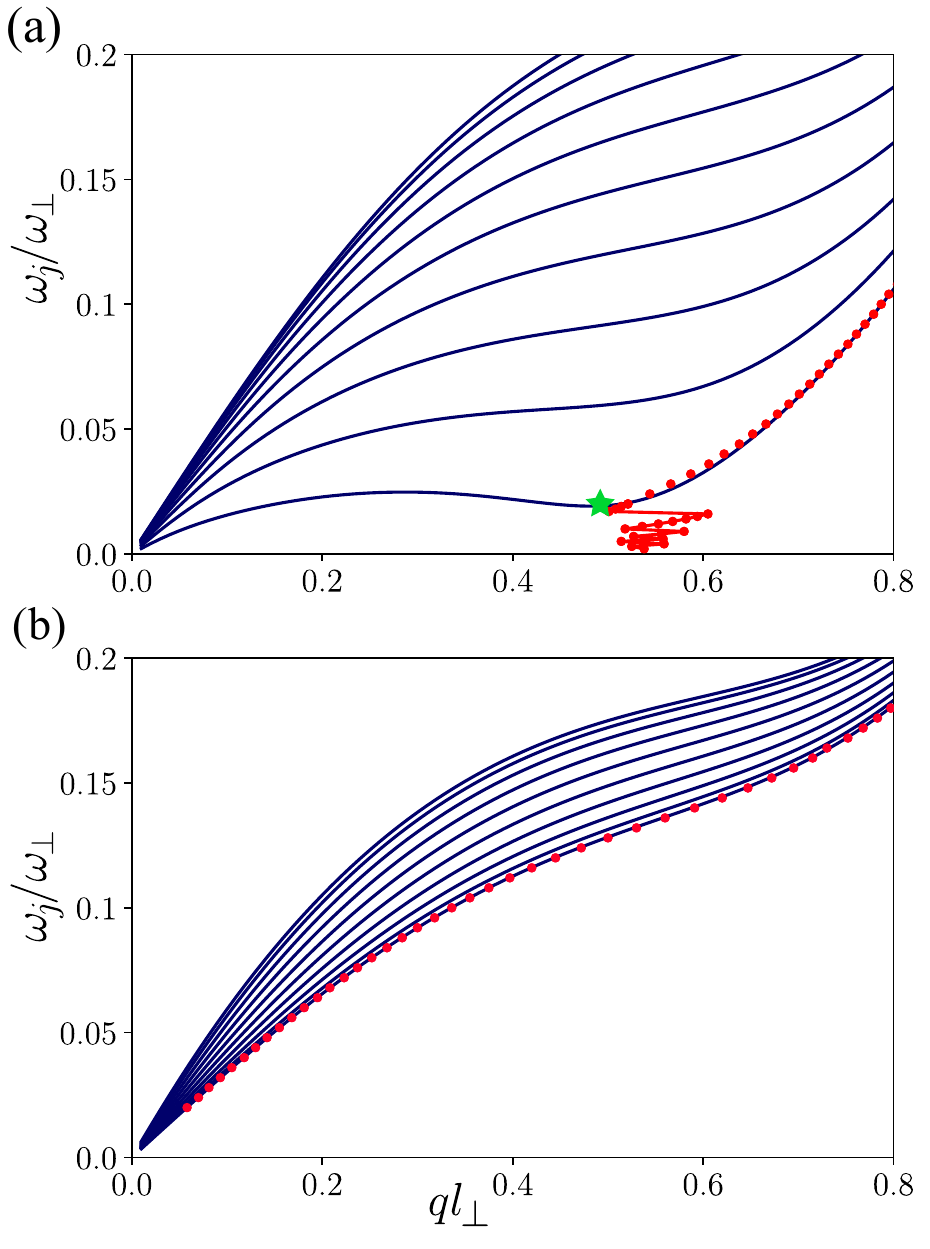}
         \caption{Bogoliubov spectrum of ten Q1D homogeneous dipolar condensates with $gn_0 = -0.06\hbar\omega_\perp$, $g_d n_0 = 0.14\hbar\omega_\perp$, and $\Delta = 5l_\perp$  with the dipole orientation (a) $\phi =0$ degrees and (b) $\phi =90$ degrees. Filled circles (red) show the most unstable momentum for the amplitude of modulation $\alpha=0.06$ as an increasing function of driving frequency $\omega_m$ (refer Sec.~\ref{pd}), which follows the lowest branch. In (a), the roton minimum (marked by a star) causes a nontrivial wavenumber selection when $\omega_m<\omega_r$, where $\omega_r$ is the roton mode frequency. In particular, the higher harmonics such that $n\omega_m\gtrapprox\omega_r$ with $n=2, 3, ...$ become more unstable, putting a lower bound to the unstable momenta appearing in the system  \cite{nat10, lako12, tur20}.}
        \label{fig:2}
\end{figure}

Considering the following solution $\psi_j(z,t)=\left[\sqrt{n_0}+u_je^{i(qz - \omega t)}+v_j^{\ast}e^{-i(qz-\omega t)}\right]e^{-i\mu_jt/\hbar}$ in Eq.~(\ref{gpe}), where $n_0$ is the homogeneous density and up to linear order in the amplitudes $u_j$ and $v_j$, we obtain the coupled Bogoliubov-de Gennes equations,
\begin{eqnarray} 
\label{bdge1}
     \left(E_q+gn_0\right)u_j+gn_0v_j+\frac{g_dn_0}{3}\sum_{l=1}^N F_{|l-j|}(q)[u_l + v_l]=\hbar\omega_j u_j \\
    \left(E_q+gn_0\right)v_j^{\ast}+gn_0u_j^{\ast}+\frac{g_dn_0}{3}\sum_{l=1}^N F_{|l-j|}(q)[u_l^{\ast} + v_l^{\ast}]=-\hbar\omega_j v_j^{\ast} 
        \label{bdge2}
\end{eqnarray}
where $E_q=\hbar^2q^2/2M$. Solving Eqs.~(\ref{bdge1}) and (\ref{bdge2}), we obtain the Bogoliubov spectrum of a stack of Q1D homogeneous dipolar condensates. The long-range DDIs lead to collective modes among the condensates, with their properties depending critically on the dipole orientation (see Fig.~\ref{fig:2}). For instance, when $\phi=0$ degrees, the inter-tube DDIs are attractive, and the lowest excitation branch corresponds to in-phase density oscillations among neighboring condensates. The highest mode represents the out-of-phase density modulations. In contrast, for $\phi=90$ degrees, due to the repulsive inter-tube DDIs, the lowest excitation mode represents out-of-phase density modulations, and the highest one characterizes in-phase density modulations among the neighboring condensates. As a result, we have a sparse spectrum for $\phi=0$ degrees [see Fig.~\ref{fig:2}(a)] and a dense spectrum for $\phi=90$ degrees [see Fig.~\ref{fig:2}(b)]. Particularly, note that we can engineer the nature of the lowest excitation branch by tuning the dipole angle $\phi$. Since we have taken $g<0$, a roton-like minimum appears in the lowest mode for $\phi=0$ degrees as shown in Fig.~\ref{fig:2}(a) \cite{lako12,nat10}. Henceforth, we restrict the interaction parameters such that the Bogoliubov spectrum is stable (purely real), and the Lee–Huang–Yang (LHY) quantum correction to the chemical potential from quantum fluctuations can be neglected \cite{edl17,pri21}.


\section{Parametric driving}
\label{pd}

In the following, we consider the parametric driving of the $s$-wave scattering length, $a_s(t) =  \bar a_s\left[1+2\alpha\cos(2\omega_m t)\right]$ with amplitude $\alpha$ and frequency $\omega_m$. Consequently, the array of homogeneous Q1D condensates becomes unstable against forming Faraday patterns. To understand the wavenumber selection of the patterns, we introduce $\psi_j(z,t) =\psi_j^{H}(z,t)\left[1+K_j(t)\cos qz \right]$ in Eqs.~(\ref{gpe}), where
 \begin{equation}
 \psi_j^{H}(z,t) = \sqrt{n_0} \exp\left(-i\left[\frac{\mu_j t}{\hbar} + \frac{\alpha\bar gn_0}{\hbar\omega_j}\sin{2\omega_mt}\right]\right)
 \end{equation}
 is the homogeneous solution in the presence of periodic modulation with $\bar g=2\hbar^2\bar a_s/ml_\perp^2$, and $K_j(t)=r_j(t)+is_j(t)$ is the amplitude of the density modulation. Linearizing in $K_j(t)$, we arrive at two first order coupled differential equations for $r(t)$ and $s(t)$. The latter can be combined into a single second order equation as,
\begin{eqnarray} 
\label{ces}
\hbar^2\frac{d^2r_j}{dt^2} + E_q\left[ E_q + 2g(t) n_0 \right]r_j + \frac{2}{3} g_d E_q \sum_{l=1}^NF_{|l-j|}(q) r_l = 0, 
\end{eqnarray}
which can be rewritten as
\begin{eqnarray}
    \hbar^2\frac{d^2\bm R(t)}{dt^2} +\mathcal{A}\bm R(t) = 0,
    \label{Re}
\end{eqnarray}
where $\mathcal{A}=E_q\left[ E_q + 2g(t)n_0 \right]\mathbb{I}+\mathcal{G}(q)$, $\mathbb{I}$ is an $ N\times N $ identity matrix and $\bm R = (r_1,r_2,\dots r_N)^T$. The matrix $\mathcal G(q)$ is given by,
\begin{eqnarray} \label{eq:G_matrix}
    \mathcal G(q) = \dfrac{2}{3} g_d E_q\begin{pmatrix}
           F_{0}(q) &  F_{1}(q) & F_{2}(q) & \dots  \\ 
          F_{1}(q) & F_{0}(q) & F_{1}(q) & \dots \\
           F_{2}(q) &  F_{1}(q) & F_{0}(q) & \dots \\
           \vdots & \vdots & \vdots & \ddots
    \end{pmatrix},
\end{eqnarray}
where $F_{j}(q)$ is given in Eq.~(\ref{fn}). Equation~(\ref{Re}) is finally decoupled into $N$ independent Mathieu-like equations:
\begin{eqnarray}
\hbar^2\frac{d^2\bar{r}_j}{dt^2} + \left[ \epsilon^2_j(q)+4\alpha\bar g_0 n_0E_q\cos 2\omega_m t \right]\bar{r}_j =0,
\label{mes}
\end{eqnarray}
where $\bar{r}_j=\bm g_j^T\bm R$ is the collective modulation amplitude of the stack of dipolar condensates with $\bm g_j$ being the $j$th eigenvector of $\mathcal{A}$ and $\epsilon_j(q)=\hbar\omega_j$ is the Bogoliubov dispersion. The solutions of Eq.~(\ref{mes}) are of the form $\bar r_j(t)= f_j(t)e^{\sigma_j t} $, where $\sigma_j$ is the Floquet exponent and $f_j(t)=f_j(t+2\pi/\omega_m)$.  ${\rm Re}(\sigma_j)>0$ indicates the dynamical instability of the stack of condensates against the formation of transient density patterns. In the limit $\alpha\to 0$, the unstable momenta are determined by the resonances $n\hbar\omega_m=\epsilon_j(q)$, $\forall j$. The most unstable momentum governs the wavelength of the transient density pattern, i.e., one with the largest $\rm{Re}(\sigma_j)$. It is further verified by numerically solving the coupled NLGPEs (\ref{gpe}) starting from a homogeneous solution with a tiny random noise embedded in it. 

\subsection{Stripe pattern}
\label{sp}

\begin{figure}
\centering
    \includegraphics[width=\linewidth]{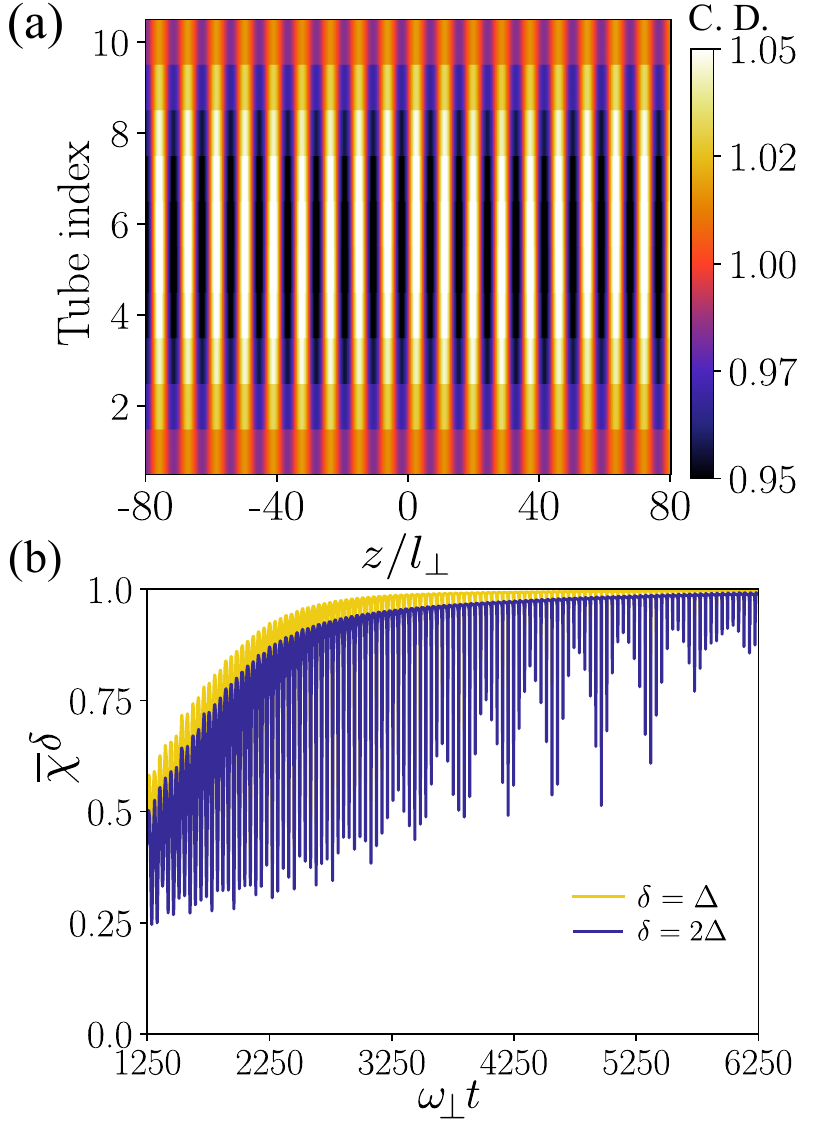}
    \caption{(a) Stripe density pattern and (b) the corresponding density-density correlations in a stack of ten Q1D dipolar condensates for $\phi=0$ degrees, $g n_0 = -0.06\hbar\omega_\perp$, $ g_dn_0 = 0.14\hbar\omega_\perp $, $ \Delta = 5l_\perp$, $\alpha = 0.01$ and $\omega_m=0.07\omega_\perp$. The snapshot of the pattern is taken at $\omega_\perp t = 6185$. The average density-density correlation $\bar{\chi}^\delta$ is shown for nearest ($\delta=\Delta$) and next-nearest condensates ($\delta=2\Delta$). While plotting, we have taken a finite width of the condensate along the transverse $y$-direction for better visualization. C.D. stands for condensate density.} 
    \label{fig:3}
\end{figure}

First, we consider dipoles oriented along the $y$-axis ($\phi=0$ degrees). In this case, the inter-tube DDIs are attractive, and the lowest mode in the Bogoliubov spectrum characterizes in-phase density modulations among neighboring condensates. In particular, we consider the case of Fig.~\ref{fig:2}(a) in which the lowest branch has a roton minimum with a frequency $\omega_r$ and momentum $k_r$ [marked by a star in Fig.~\ref{fig:2}(a)]. Upon periodic driving, the solutions of the Eq.~(\ref{mes}) reveal that the unstable momenta are provided by the resonances $n\hbar\omega_m=\epsilon_j(q)$, which are $N$ in the count for a given driving frequency and $n$. Since the Floquet exponent $\sigma_j(q)$ is proportional to $q^2$ \cite{nat10}, the highest among the $N$ momenta, which is that of the lowest mode, is the most unstable. The excitation of the lowest mode leads to a stripe density pattern, as shown in Fig.~\ref{fig:3}(a), where the numerical results are shown for ten Q1D condensates. Due to the finite number of tubes, the lowest mode has a larger amplitude for the central condensate, decreasing as we move to the outer ones. As a consequence, the contrast of the pattern is also minimal in the outer condensates. For $\omega_m>\omega_r$, it is the first harmonics (n=1), which is most unstable, whereas for $\omega_m<\omega_r$, the higher harmonics having energy greater and closer to the roton minimum become more unstable, i.e., $n\omega_m\gtrapprox\omega_r$ with $n=2, 3, ...$. \cite{nat10, lako12, tur20}. The latter leads to a non-monotonous wavenumber selection, as shown in Fig.~\ref{fig:2}(a)].

Further, to characterize the pattern, we calculate the density-density correlation function between a pair of condensates,
  \begin{equation} \label{eq:corr_func}
    \chi^{(j,k)}(t) = \frac{\int dz~S_{j}(z,t) S_{k}(z,t)}{\sqrt{\left(\int dz~S_{j}^2(z,t)\right)} \sqrt{\left(\int dz~S_{k}^2(z,t)\right)}}
\end{equation}
where $S_{j}(z,t) = n_{j}(z,t) - n_0$ with $j$ denoting the tube index along the $y$-axis. Such correlations can be extracted from in situ imaging of density of all the tubes, capturing simultaneously, as detailed in \cite{hun11, her21}. The average density-density correlation among all pairs of tubes separated by a distance  $\delta$ is defined as,
\begin{equation} \label{eq:corr_func_delta}
    \bar\chi^\delta(t) = \frac{1}{N-\delta/\Delta}\sum_{j=1}^{N-\delta/\Delta} \chi^{(j,j+\delta/\Delta)}(t).
\end{equation}
In Fig.~\ref{fig:3}(b), we show $\bar\chi^\delta(t)$ of the stripe pattern for neighbouring ($\delta=\Delta$) and nearest-neighbouring ($\delta=2\Delta$) condensates. Since the density modulations are in phase, $\bar\chi^\delta$ is positive irrespective of $\delta$. Since the amplitude of the density pattern oscillates with the driving frequency, $\bar\chi^\delta(t)$ also exhibits oscillations of frequency $\omega_m$. Eventually,  $\bar\chi^{\delta=\Delta}(t)$ attains a maximum value of $\sim 1$, indicating a maximally correlated density pattern.


\subsection{Checkerboard pattern}
\label{cbp}

\begin{figure}
\centering
    \includegraphics[width=\linewidth]{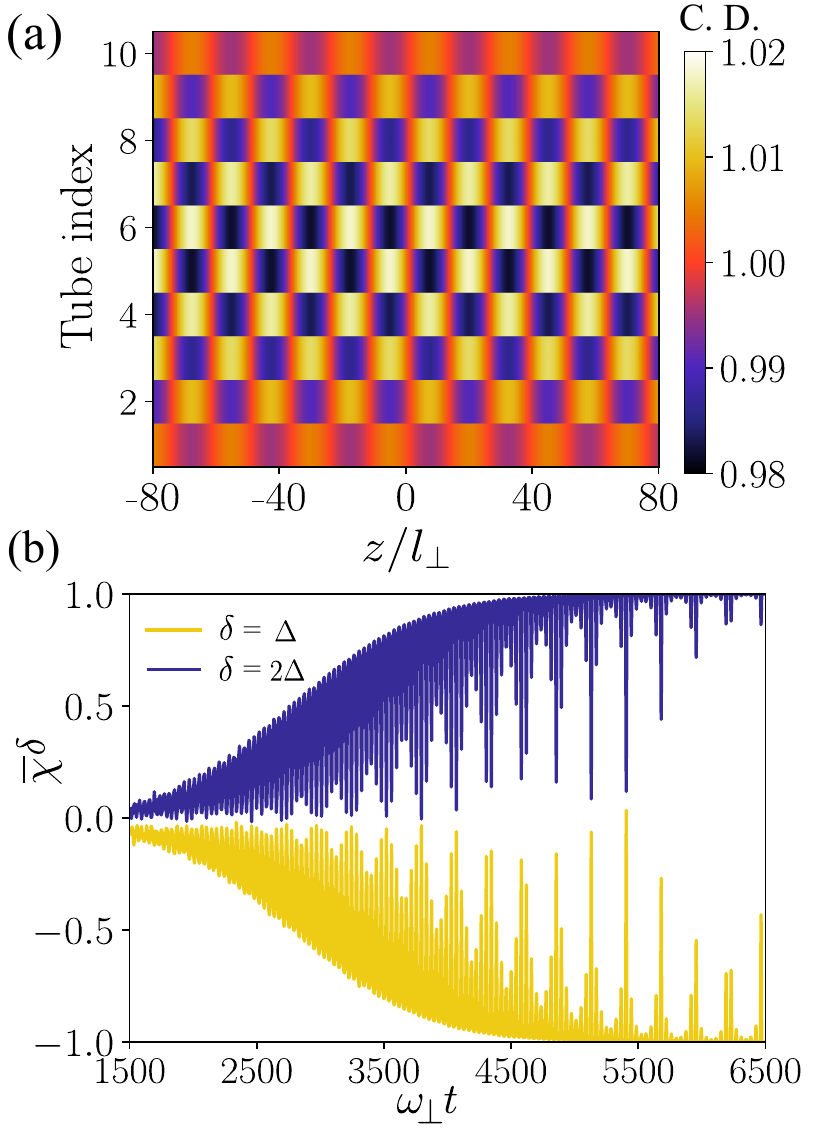}
    \caption{(a) Checkerboard density pattern and (b) the corresponding density-density correlations in a stack of ten Q1D dipolar condensates for $\phi =90$ degrees, $g n_0 = -0.06\hbar\omega_\perp$, $ g_dn_0 = 0.14\hbar\omega_\perp $, $ \Delta = 5l_\perp$, $\alpha = 0.06$ and $\omega_m=0.08\omega_\perp$. The snapshot of the pattern is taken at $\omega_\perp t = 6500$. While plotting, we have taken a finite width of the condensate along the transverse $y$-direction for better visualization. The average density-density correlation $\bar{\chi}^\delta$ is obtained for nearest ($\delta=\Delta$) and next-nearest condensates ($\delta=2\Delta$). C.D. stands for condensate density.} 
    \label{fig:4}
\end{figure}

Changing the dipole orientation from $y$-axis ($\phi=0$ degrees) to $x$-axis ($\phi=90$ degrees) changes the lowest mode character from in-phase to out-of-phase density modulations among the neighboring condensates. It leads to the formation of a checkerboard pattern when the $s$-wave scattering length is periodically modulated [see Fig.~\ref{fig:4}(a)]. We consider the spectrum in Fig.~\ref{fig:2}(b), and the absence of a roton minimum makes the most unstable momentum a monotonous function of the driving frequency. For the checkerboard pattern, the density modulations between the neighboring condensates are out of phase, and hence, $\bar\chi^{\delta=\Delta}$ negative. At longer times, $\bar\chi^{\delta=\Delta}$ attains a value of -1, indicating a maximally anti-correlated density modulation in neighboring condensates [see Fig.~\ref{fig:4}(b)]. The density modulations in any pair of next-nearest-neighbor condensates are in phase, making $\bar\chi^{\delta=2\Delta}(t)\sim 1$ at longer times. Thus, condensates separated by odd multiples of $\Delta$ are anti-correlated in a checkerboard pattern, and even multiples are correlated.

\section{Quenching the dipole orientation: Dynamical Pattern transitions}
\label{qda}
\begin{figure*}
      \centering
     \includegraphics[width=\linewidth]{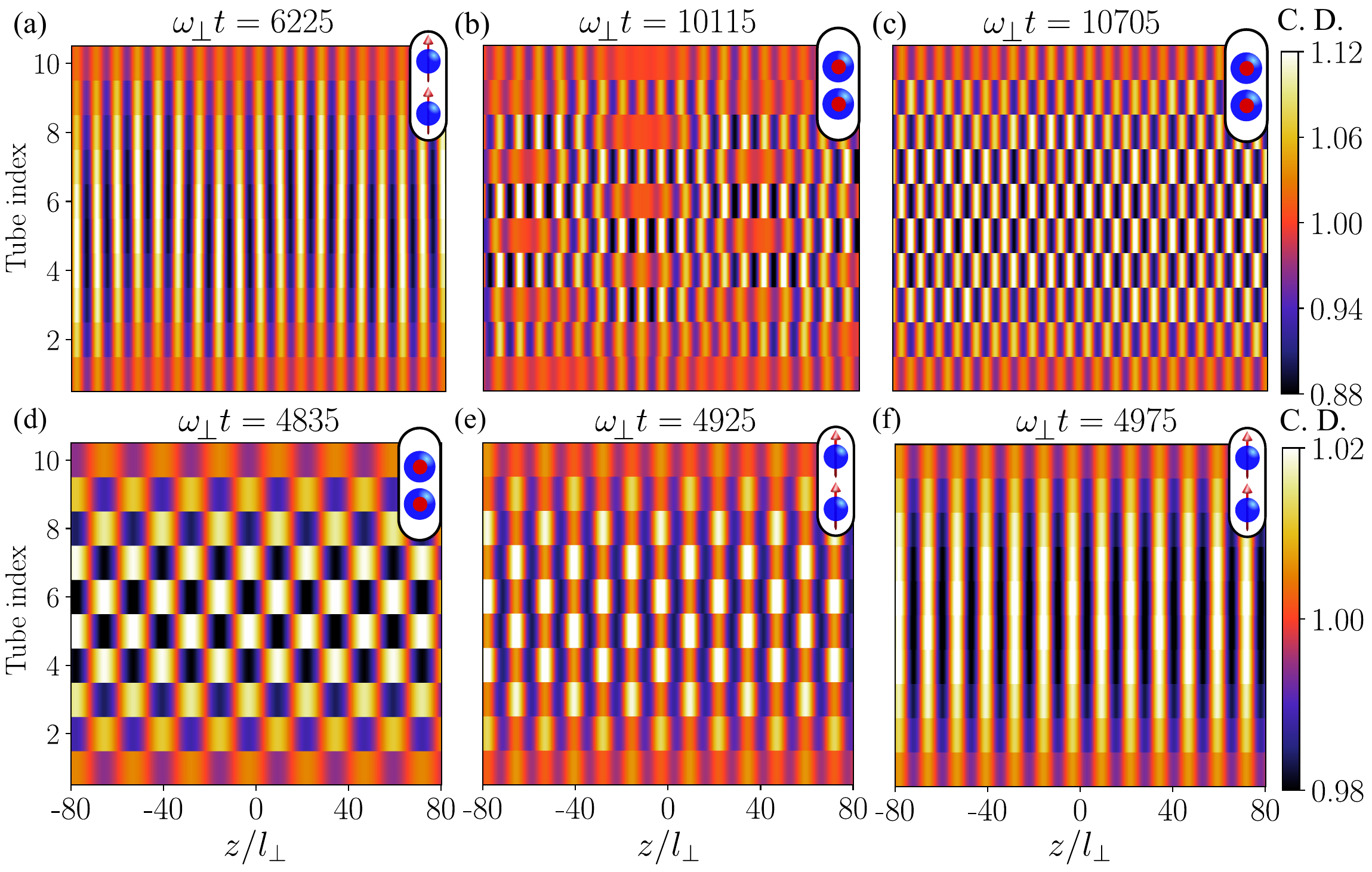}
     \caption{Dynamical pattern transition under instantaneous quench of dipole orientation. (a)-(c) shows the pattern transition when $\phi$ is quenched from zero to 90 degrees, and (d)-(f) shows the same for the reverse case. The instant of time for each density snapshot is provided at the top of each figure. For (a)-(c), $\omega_m=0.07\omega_\perp$ and $\alpha=0.01$ and we quench at $\omega_{\perp} t = 6250$.  For (d)-(f), the modulation parameters are $\omega_m = 0.08\omega_\perp$ and $\alpha=0.08$. The dipole angle $\phi$ is quenched at $\omega_\perp t = 4850$.The orientation of the dipoles is shown in each figure. The other parameters are $g n_0 = -0.06\hbar\omega_\perp$, $ g_dn_0 = 0.14\hbar\omega_\perp $, and $ \Delta = 5l_\perp$. C.D. stands for condensate density.}
     \label{fig:5}
 \end{figure*}

\begin{figure}[h!]
     \centering
     \includegraphics[width=\linewidth]{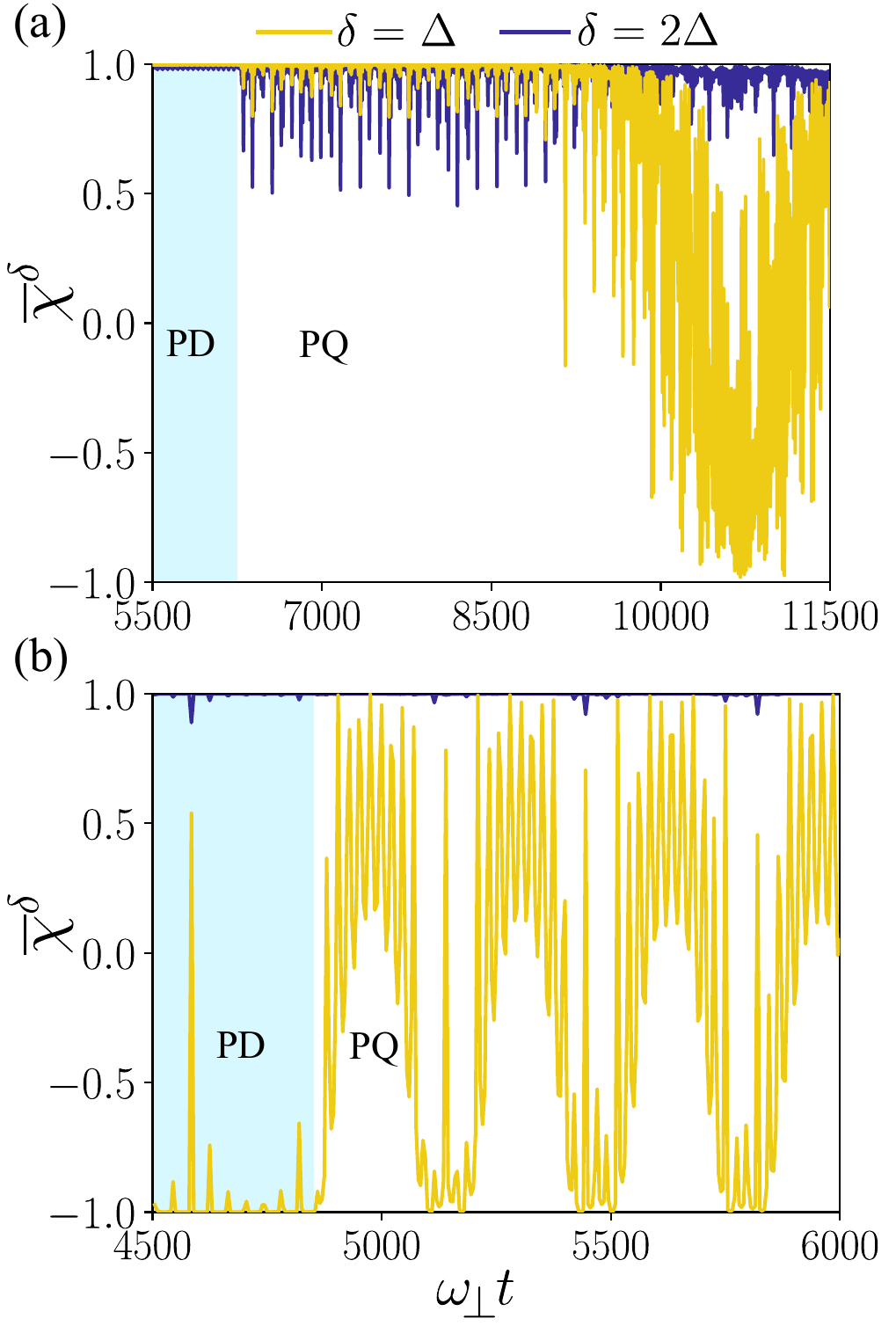}
     \caption{{\em Abrupt quench}. The evolution of density-density correlation function $\bar\chi^\delta$ for the driven-quench dynamics with (a) $\phi=0$ to $\phi=90$ degrees and (b) $\phi=90$ degree to $\phi=0$ quenches. For (a), the modulation parameters are $\omega = 0.07\omega_\perp$ and $\alpha=0.01$ and for (b), $\omega_m = 0.08\omega_\perp$ and $\alpha=0.08$. The other parameters are the same as in Fig.~\ref{fig:5}. In (a) the quench is done at $\omega_{\perp} t = 6250$ and in (b) at $\omega_{\perp} t =4850$. PD stands for periodic driving, and PQ stands for post-quench.}
     \label{fig:6}
 \end{figure}
Interestingly, once the pattern is formed via periodic driving, quenching the dipole orientation or the angle $\phi$ leads to the dynamical transition between the checkerboard and stripe patterns. For instance, when $\phi=0$ degrees initially, the periodic driving leads to a stripe pattern. Once the amplitude of the pattern reaches sufficiently high, the periodic driving is stopped and simultaneously quenches $\phi$ to 90 degrees. The quenching dynamically transits the stripe pattern to a checkerboard pattern. Similarly, starting from $\phi=90$ degrees, creating a checkerboard pattern, and then quenching $\phi$ to 0 degrees leads to checkerboard to stripe transition.

The first row of Fig.~\ref{fig:5} shows the dynamical transition of patterns due to an instantaneous quench of $\phi$ from zero to 90 degrees. First, we form the stripe pattern for $\phi=0$ via periodic modulation [see Fig.~\ref{fig:5}(a)]. Then, the driving is stopped and instantly quenches $\phi$ to 90 degrees. The quenching destabilizes the stripes [Fig.~\ref{fig:5}(b)] since the inter-tube DDIs (along the $y$-axis) abruptly changed from maximally attractive ($\phi=0$) to maximally repulsive ($\phi=90$ degrees). Subsequently, the stripes break, and the density peaks self-organize into a checkerboard pattern as shown in Fig.~\ref{fig:5}(c). Similarly, as shown in the second row of Fig.~\ref{fig:5}, the initial checkerboard pattern created by periodic driving for $\phi=90$ degrees dynamically changes to a stripe pattern upon quenching $\phi$ to zero degrees. Quenching changes the inter-tube repulsion into attractive interactions. As a result, each density peak in the checkerboard pattern breaks into two due to the attractive pull from the density peaks on either side of the neighboring tubes. Eventually, they align into stripe patterns as shown in Figs.~\ref{fig:5}(e) and ~\ref{fig:5}(f). The intermediate stripe pattern in Fig.~\ref{fig:5}(e) also exhibits a density modulation along each stripe. The breaking of density peaks makes the spatial periodicity of the stripe pattern half that of the initial checkerboard pattern.

The evolution of the average density-density correlation $\bar\chi^\delta$ during the pattern transition is shown in Fig.~\ref{fig:6}. In Fig.~\ref{fig:6}(a), the nearest neighbor correlation ($\delta=\Delta$) changes from positive to negative, whereas the next nearest neighbor correlation ($\delta=2\Delta$) remains positive, indicating the transition from a stripe to a checkerboard pattern. In contrast, for the checkerboard to stripe transition, $\bar\chi^{\delta=\Delta}$ changes its sign from negative to positive [see  Fig.~\ref{fig:6}(b)]. For a fixed $g$ and $g_d$, the maximally attractive ($\phi=0$ degrees) inter-tube DDI is two times stronger than the maximally repulsive ($\phi=90$ degrees) one, causing the checkerboard to stripe transition much quicker than the stripe to checkerboard transition. In general, the transition period between the patterns depends on several factors, particularly the pattern amplitude before the quenching. The longer the initial periodic driving (before the condensates get destroyed by heating), the larger the amplitude and the stronger the DDIs between the density peaks in the neighboring tubes. The latter can augment the speed of the pattern transition upon the abrupt quench. At the same time, periodic driving also increases the average kinetic energy. The quenching further excites the vibrational modes of the initial transient crystalline-like pattern, which can lead to subsequent dynamical oscillations between the two patterns, as evident in the dynamics of the density correlations. 

\begin{figure}[h!]
     \centering
     \includegraphics[width=\linewidth]{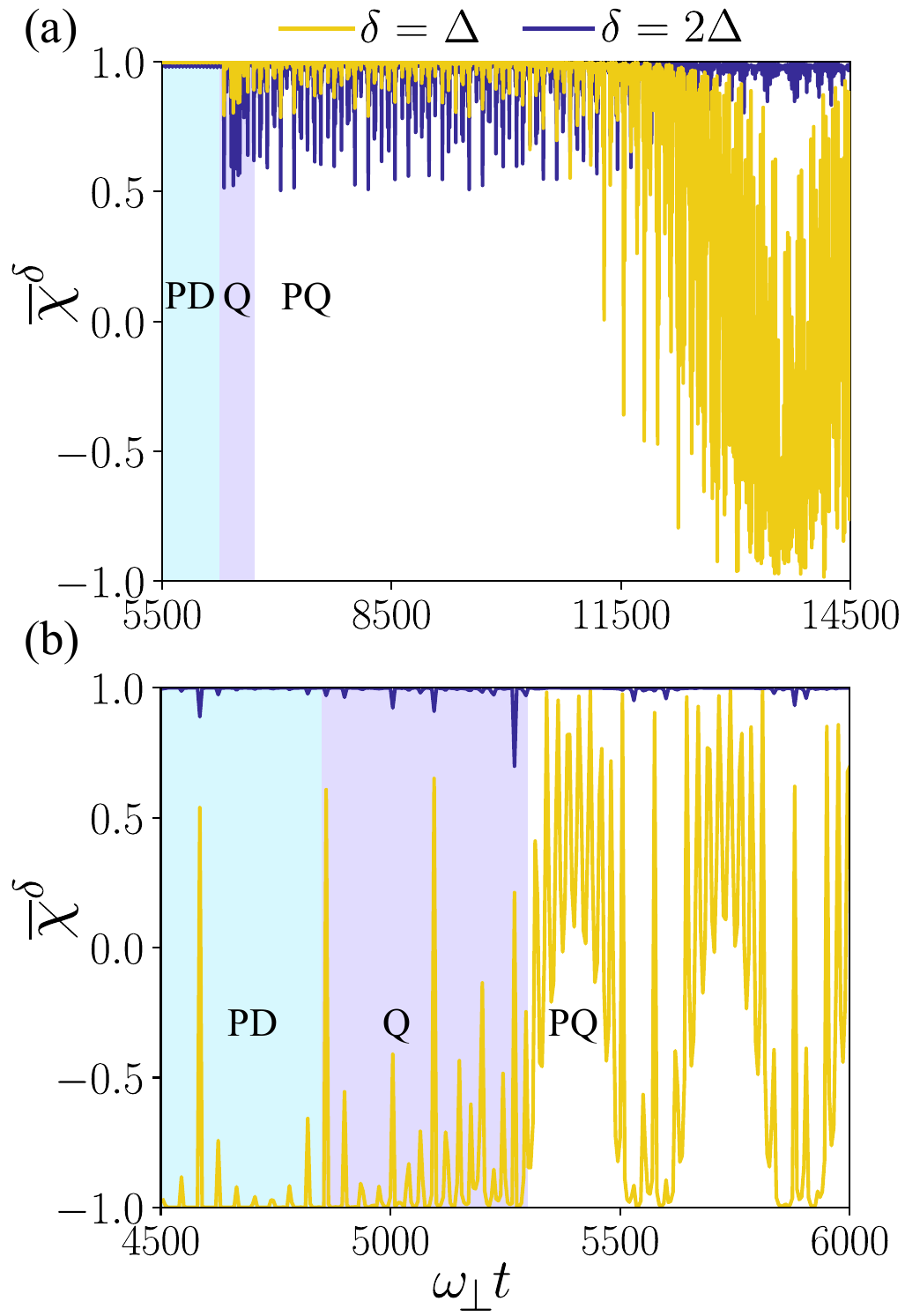}
     \caption{{\em Linear quench}. The evolution of density-density correlation function $\bar\chi^\delta$ for the driven-quench dynamics with (a) $\phi=0$ to $\phi=90$ degrees and (b) $\phi=90$ degree to $\phi=0$ quenches.  For (a), the modulation parameters are $\omega = 0.07\omega_\perp$ and $\alpha=0.01$and for (b), $\omega_m = 0.08\omega_\perp$ and $\alpha=0.08$. The other parameters are same as in Fig.~\ref{fig:5}. In (a) the quench is started at $\omega_{\perp} t = 6250$ and ends at $\omega_{\perp} t = 6695$, and in (b) from $\omega_{\perp} t = 4850$ to $\omega_{\perp} t = 5295$. PD stands for periodic driving, Q for the regime of the linear quench and PQ for post-quench.}
     \label{fig:7}
 \end{figure}

Though linear quench also results in qualitatively similar dynamics, the transition period has been prolonged and depends on the quench rate. The latter could also be analyzed via the average density correlations, as shown in Fig.~\ref{fig:7}. If quenched sufficiently slowly, excitation of the vibrational modes of the transient crystalline pattern can be suppressed. Hence, a complete revival of the initial pattern may not occur after the transition.
 

\section{Summary and outlook}
\label{sum}
Summarizing, we analyzed the formation of checkerboard and stripe patterns in a stack of driven Q1D homogeneous dipolar condensates. The parametric driving of $s$-wave scattering length excited the lowest Bogoliubov mode. The checkerboard and stripe patterns are observed by engineering the lowest mode via the dipole orientation. The dynamical transition between the two patterns is seen by either abruptly or linearly quenching the dipole orientation after the initial pattern is formed via parametric driving.

Our studies open up several exciting perspectives. One aspect would be designing driving schemes to selectively excite any mode in the Bogoliubov spectrum of a stack of disconnected dipolar condensates. Second, to investigate the nature of patterns at an arbitrary orientation of the dipoles and their response to quenching of different interaction parameters. The studies can also easily extend two-dimensional multi-layer dipolar BECs, where the coupled patterns can be more complex. Note that the Lieb-Liniger parameter of the system explored here is much smaller than one, justifying mean-field calculations. Recently, there has been growing interest in exploring one-dimensional Bose systems in the Tonks-Girardeau regime, with dipolar atoms and molecules, near the boundary of integrability \cite{tan18, kao21} and also deep into the
non-integrable regime. It would be interesting to explore the response of dipolar gases on the periodic modulation of $s$-wave scattering length in these regimes.


\section{Acknowledgements}
We acknowledge National Supercomputing Mission (NSM) for providing computing resources of "PARAM Brahma" at IISER Pune, which is implemented by C-DAC and supported by the Ministry of Electronics and Information Technology (MeitY) and Department of Science and Technology (DST), Government of India. R.N. further acknowledges DST-SERB for Swarnajayanti fellowship File No. SB/SJF/2020-21/19, and the MATRICS grant (MTR/2022/000454) from SERB, Government of India. S.N. acknowledges funding from DST through KVPY scholarship.

\bibliographystyle{apsrev4-1}
\bibliography{lib.bib}
\end{document}